\documentclass[conference]{jpaper}

\newcommand{\shan}[1]{{\color{red}(shan: #1)}}

\newcommand{\myquote}[1]{
\begin{quote}
\centering
\small
\textit{#1}
\end{quote}
}
\newcommand{\jira}[3]{\href{http://issues.apache.org/jira/browse/#1-#3}{#2-#3}}
\newcommand{\ca}[1]{\jira{CASSANDRA}{CASSANDRA}{#1}}
\newcommand{\hb}[1]{\jira{HBASE}{HBASE}{#1}}
\newcommand{\hd}[1]{\jira{HDFS}{HDFS}{#1}}

\usepackage[numbers,sort]{natbib}

\usepackage[normalem]{ulem}
\usepackage{listings}
\usepackage[table]{xcolor}
\usepackage{pifont}
\usepackage{color,soul}
\usepackage{colortbl}
\usepackage{subfig}
\usepackage{amsfonts}
\usepackage{xspace}
\usepackage{tabularx}

\usepackage[scaled=.7]{beramono}
\usepackage{flushend}
\usepackage{etoolbox}
\makeatletter
\patchcmd{\ttlh@hang}{\parindent\z@}{\parindent\z@\leavevmode}{}{}
\patchcmd{\ttlh@hang}{\noindent}{}{}{}
\makeatother

\begin{document}
\newcommand{\tool}{\textit{SmartConf}\xspace}
\newcommand{\CPbug}{Configuration Related Performance Issues\xspace}
\newcommand{\cpbug}{CP Issues\xspace}
\newcommand{\seca}{Sec.\xspace}
\newcommand{\cp}{control theory\xspace}
\newcommand{\etal}{\textit{et al}.} 

\newcommand{\wrt}{\textit{w.r.t} } 

\title{Understanding and Auto-Adjusting Performance-Related Configurations}

\author{
 {\rm Shu Wang \quad Chi Li \quad $^\dagger$William Sentosa \quad Henry Hoffmann \quad Shan Lu} \\ University of Chicago \quad $^\dagger$Bandung Institute of Technology
 \\ \{shuwang, lichi\}@uchicago.edu \quad williamsentosa@students.itb.ac.id
 \\ hankhoffmann@cs.uchicago.edu \quad shanlu@uchicago.edu
}

\date{}
\maketitle

\thispagestyle{empty}

\begin{abstract}
Modern software systems are often equipped with hundreds to thousands of configuration options, many of which greatly affect performance. Unfortunately, properly setting these configurations is challenging for developers due to the complex and dynamic nature of system workload and environment.
In this paper, we first conduct an empirical study to understand performance-related configurations and the challenges of setting them in the real-world. Guided by our study, we design a systematic and general control-theoretic framework, \tool, to automatically set and dynamically adjust performance-related configurations to meet required operating constraints while optimizing other performance metrics. Evaluation shows that \tool is effective in solving real-world configuration problems, often providing better performance than even the best static configuration developers can  choose under existing configuration systems.
\end{abstract}

\section{Introduction}
\label{sec:intro}

\myquote{``all constants should be configurable, even if we can't see any reason to configure them.'' ---
 \hd{4304}}

\subsection{Motivation}
\vspace{-2mm}
Modern software systems are equipped with hundreds to thousands of configuration options allowing customization to different workloads and hardware platforms.  While these configurations provide great flexibility, they also put great burden on users and developers, who are now responsible for setting them to ensure the software is performant and available.  Unfortunately, this burden is more than most users can handle, making software \emph{misconfiguration} one of the biggest causes of system misbehavior
\cite{fragile,Gray86heisenbug,tianyin.sosp13}.
Misconfiguration leads to both incorrect functionality 
(e.g., wrong outputs, crashes) and poor performance. Although recent research has tackled functionality issues arising from misconfiguration \cite{tianyin.sosp13,pcheck.osdi16}, poor performance is an open problem.


In server applications, customizable configuration parameters are especially common. 
These configurations control the size of critical data structures, 
the frequency of performance-related operations, the thresholds and weights in workload-dependent algorithms, and many other aspects of system operation. 
Previous studies find that 20\% of user-reported misconfiguration problems result in severe performance degradation, and yet, performance-related misconfigurations are under-reported \cite{yin2011empirical}.  
Additional surveys show about a third of Hadoop's misconfiguration problems result in \texttt{OutOfMemoryError}s 
\cite{rabkin2013hadoop}.

Setting performance-related configurations, \emph{PerfConfs} for short, is challenging because they represent tradeoffs; e.g., between memory usage and response time.  Managing these tradeoffs requires deep knowledge of the underlying hardware, the workload, and the PerfConf itself. Often, these relationships are not, or cannot, be clearly explained in the documentation \cite{howtorationalize}.  Even with clear documentation, the workload and system interaction are often too \textit{complicated} or \textit{change} too
quickly for users to maintain a proper setting \cite{rabkin2013hadoop}.
In many cases, there is simply no satisfactory static setting \cite{itask.sosp15}.

Configuring software to an optimal point in a tradeoff space is a constrained optimization problem.  Operating requirements represent constraints, and the goal is finding the optimal PerfConf setting given those constraints.  For example, a larger queue makes a system more responsive to bursty requests at the cost of increased memory usage.  Here the constraint is that the system not run out of memory and the goal is to minimize response time. Prior work addresses this problem in several ways, with no perfect solution.

The industry standard is simply to expose parameters to users who are forced to become both application and system experts to understand the best settings for their particular system and workload, as shown in Section \ref{sec:study}.
Machine learning techniques explore complex configuration spaces to find near optimal settings \cite{morpheus.osdi16,pavlo.sigmod17,starfish.cidr11}; however, they are less well suited for managing constraints, especially in dynamic environments \cite{TAAS}. Control theoretic frameworks handle constrained optimization for non-functional software properties \cite{ControlSurvey}. Typical control solutions, however, require deep understanding of a specific application or system (e.g., \cite{Horvarth,LuEtAl-2006a,SunDaiPan-2008a,Agilos,GRACE}), and hence, are impractical for real-world developers to adopt. Even  general control synthesis techniques (e.g., \cite{filieri2014automated}) still require user-specified parameters.  More importantly,
they cannot handle challenges unique to PerfConfs, such as hard constraints---e.g., not going out of memory---and indirect relationships between PerfConfs and performance. 

\subsection{Contributions}
\vspace{-2mm}
In this paper, we first conduct an empirical study to understand real-world performance-related configuration problems. The study's results motivate us to construct a general framework, \tool.
Unlike traditional configuration frameworks---where users set PerfConfs at system launch---\tool automatically sets and dynamic adjusts PerfConfs. \tool decomposes the PerfConf setting problem to let users, developers, and control-theoretic techniques---which we specifically design for PerfConfs---each focus on what they know the best, as shown in Table \ref{tab:framework}.


\begin{table}
\begin{tabular}{l|l|l}
\textbf{Prior} & \textbf{Who answers these questions?}& {\textbf \tool}\\
\toprule
n/a      &Which C needs dynamic adjustment?& {\footnotesize Developers}\\
n/a      &What perf. metric M does C affect? & {\footnotesize Developers}\\
n/a      &What is the constraint on metric M? & {\footnotesize Users}\\
{\footnotesize Users}       &How to set \& adjust configuration C? & {\footnotesize \tool} \\
\end{tabular}
\caption{Traditional configuration vs \tool configuration}
\label{tab:framework}
\end{table}

\paragraph{Empirical study} We look at
80 developer-patches and 54 user-posts concerning PerfConfs
in 4 widely used large-scale systems.
We find that (1) PerfConfs are common among configuration-related patches ($>$50\%) and forum questions ($\sim$30\%);
(2) almost half of PerfConf patches fix performance problems caused by improper default settings;
(3) properly setting most PerfConfs requires considering dynamic workload, environmental factors, and performance tradeoffs. 

Our study also points out challenges in setting and adjusting PerfConfs: (1) about half of PerfConfs threaten \textit{hard} performance constraints like out-of-memory or out-of-disk problems; (2) about half of PerfConfs affect performance \textit{indirectly} through setting thresholds for other system variables; (3) more than half of PerfConfs are associated with specific system events and hence only take effect \textit{conditionally}; and (4) often different configurations affect the same performance goal simultaneously, requiring \textit{coordination}.

\paragraph{\tool interface} Guided by this study, we design a new configuration interface.
For developers, \tool encourages
them to decide which PerfConf should be dynamically configured and enables them to easily convert a wide variety of PerfConfs from their traditional format---requiring developers/users to set manually at application launch---into an automatically adjustable format.
For users, \tool allows them to specify the performance constraints they desire, without worrying about how to set and adjust PerfConfs to meet those constraints while optimizing other performance metrics.

\paragraph{\tool control-theoretic solution} 
To automate PerfConf setting and adjustment, we explore a systematic and general control-theoretic solution to implement \tool library.  We explicitly design for the four PerfConf challenges noted above, without introducing any extra parameter tuning tasks for developers or users---problems that were \textbf{not} handled by existing control theoretic solutions. 

\paragraph{Evaluation} Finally, we apply the \tool library to solve real-world PerfConf problems
in widely used open-source distributed systems (Cassandra, HBase, HDFS, and MapReduce). 
With only 8--76 lines of code changes, we easily refactor a problematic configuration to automatically adjust itself and 
deliver
better performance than even the best launch-time configuration settings.
Our evaluation shows that, although not a panacea, \tool framework solves many PerfConf problems
in real-world server applications.

\section{Understanding Perf-Related Configurations}
\label{sec:study} 

\myquote{ ``This is hard to configure, hard to understand, and badly documented.''
--- \hb{13919}
}


\subsection{Methodology}
\vspace{-2mm}
We study 
Cassandra (CA), HBase (HB), HDFS (HD), and Hadoop MapReduce (MR).
CA and HB are distributed 
key-value stores, 
HD is a distributed file system, 
and MR is a distributed computing infrastructure. 
They provide a good representation of modern open-source 
widely used large systems.

We first study software issue-tracking systems. The detailed developer discussion there helps us understand how and why developers introduce and change PerfConfs, as well as the trade-offs.
We first search fixed issues with key word ``config'' or with configuration files 
(e.g., \textit{hdfs-default.xml} in HD) in patches.
We then randomly sample them and manually check to see if an issue is clearly explained, about configuration, and related to performance (i.e., whether developers mentioned performance impact and made changes accordingly). We keep doing this until we find 20, 30, 20, 10 PerfConf issues for CA, HB, HD, and MR, matching the different sizes of their issue-tracking systems. 
The details are shown in Table \ref{tab:issues}.

\begin{table}
\centering
{\footnotesize
\begin{tabular}{lrrrr}
	\toprule
    & \multicolumn{2}{c}{PerfConf} &\multicolumn{2}{c}{AllConf}\\
    \cline{2-5}
	& Issues & Posts & Issues & Posts\\
	\midrule
	Cassandra&20	&20 &32&60	\\
	HBase	 &30	&7 &48&33	\\
	HDFS	 &20	&7 &31&39	\\
	MapReduce&10	&20 &13&25	\\
	\midrule
	Total	 &80	&54&124&157 \\
	\bottomrule
\end{tabular}
}
\caption{Empirical study suite}
\label{tab:issues}
\end{table}

We also search StackOverflow \cite{stackoverflow} with key words like ``config'' to randomly sample 200--300 posts for each system. We then manually read through 1000 total posts to identify configuration and PerfConf posts shown in Table \ref{tab:issues}. 
We find the StackOverflow information less accurate than the issue-trackers, and hence only discuss user complaints in Section \ref{sec:study_common}, skipping in-depth categorization.


\noindent \textbf{Threats to Validity}
This study reflects our best effort to understand PerfConfs in modern large-scale systems.
Our current study only looks at distributed systems.
We also exclude issues or posts that contain little information or are not confirmed (answered) by
developers (forum users). 
Every issue studied was cross-checked by at least two authors, and we emphasize trends that are consistent across applications. 

\subsection{Findings}

\subsubsection{How Common are PerfConf Problems?}
\label{sec:study_common}
As shown in Table \ref{tab:issues}, 65\% of issues and 35\% of posts that we studied involve performance concerns. 
 

\noindent \textbf{What are PerfConf Issues?}
For about half of the issues, either the default (24 of 80 cases)
or the original hard-coded (14 of 80 cases) setting  
caused severe performance
problems. Thus, the patch either changed a default setting
or made a hard-coded parameter configurable.
The other half simply added PerfConfs to support new features, as shown in Table \ref{tab:issuecategory}. 

\noindent \textbf{What are PerfConf Posts?}
In about 40\% of studied posts, users simply do not understand how to set a PerfConf. In another 60\%, users ask for help to improve performance or avoid out-of-memory (OOM) problems. In about half the cases, the users ask about a specific PerfConf. In other cases, the users ask whether there are any configurations they can tune to solve a performance problem, and the answers point out some PerfConfs. 
Similar to a prior study \cite{rabkin2013hadoop}, we found many posts related to OOM ($\sim$30\%).


\begin{table}
	\centering
    \footnotesize{
    \begin{tabular}{lrrrr}
    \toprule
    Category & CA & HB &  HD & MR  \\
    \midrule
    \rowcolor{gray!50}
    \multicolumn{5}{c}{Add a new configuration to ...}\\
    Tune a new functionality     &11 &16 &8 &4  \\ 
    Replace hard-coded data      &2 &1 &7 &4  \\
    Refine an existing conf.     &2 &0 &0 &1  \\ 

    \rowcolor{gray!50}
    \multicolumn{5}{c}{Change an existing configuration to ..}\\
    Fix a poor default value       &5 &13 &5 &1 \\ 
    \bottomrule
    \end{tabular}
    }
    \caption{Different types of PerfConf patches}
    \label{tab:issuecategory}
\end{table}

\vskip .5em
\subsubsection{What are PerfConfs' Impact?} ~
\label{sec:study_impact} 

\noindent \textbf{What Type of Performance do They Affect?}
As shown in Table \ref{tab:metricscount}, most PerfConfs affect user request latency.
They also commonly affect memory or disk usage, threatening server availability through OOM/OOD failures (half the cases). 
Naturally, one metric could be affected by multiple PerfConfs simultaneously, with several coordination issues \cite{mr6143,CA1007}.

As Table \ref{tab:metricscount} indicates, most PerfConfs affect multiple performance metrics (61 out of 80 issues). There are also 13 cases where the PerfConf has a trade-off between functionality and performance. For example, larger \texttt{mapreduce.job.counters.limit} provides users with more job statistics (functionality), but increases memory consumption (performance) and may even lead to OOM.

Most issue reports do not quantify performance impact. As our evaluation will show (Section \ref{sec:eval}), the impact could be huge, causing severe slow-downs or 
OOM/OOD failures.

\begin{table}
	\centering
    \footnotesize{
    \begin{tabular}{lrrrr}
    \toprule
	     & CA &HB&HD&MR \\ 
	    \toprule
User-Request Latency & 14 & 28 & 20 & 9 \\
	   
Internal Job Throughput &8 &3 &5 &0 \\
Memory/Disk Consumption &9 &15 &8 &7 \\
\midrule
\midrule
Always-on Impact        &9 &17 &8 &6 \\
Conditional Impact      &11 &13 &12 &4 \\
\midrule
\midrule
Direct Impact           &7 &16 &8 &4 \\
Indirect Impact         &13 &14 &12 &6 \\
	    \bottomrule
    \end{tabular}
    }
    \caption{How a PerfConf affects performance (one PerfConf can affect more than one metric)}
    \label{tab:metricscount}
\end{table}


\noindent \textbf{When \& How to Affect Performance?}
About half of PerfConfs affect corresponding performance metrics conditionally, being associated with a particular event or command. For example, in HDFS, \texttt{shortcircuit.streams.cache.size} decides an in-memory cache size, and affects memory usage all the time, while the number of balancing threads \texttt{balancer.moverThreads} affects user requests only during load balancing.

Almost half of the configurations directly affect performance, such as the \texttt{cache.size} and \texttt{moverThreads} mentioned above. The other half affects performance indirectly by imposing thresholds on some system variables---e.g., queue size \texttt{ipc.server.max.queue.size}, number of operations per log file \texttt{dfs.namenode.max.op.size}, and number of outstanding packets \texttt{dfs.max.packets}---which, in turn, affect performance.

\vskip .5em
\subsubsection{How to Set PerfConfs?}
\label{sec:study_set}
~

\begin{table}
	\centering
    {\footnotesize
    \begin{tabular}{lrrrr}
    \toprule
	     & CA &HB&HD&MR \\ 
	    \toprule
        \rowcolor{gray!50}
      \multicolumn{5}{c}{Configuration Variable Type}\\
Integer 	  &15 &23 &19 &9 \\ 
Floating Points &4 &5 &0 &0 \\
Non-Numerical &1 &2 &1 &1 \\
	    \midrule
        \rowcolor{gray!50}
	    \multicolumn{5}{c}{Deciding Factors}\\
Static system settings   &0 &1 &0 &1 \\
Static workload characteristics &4 &0 &0 &2 \\
Dynamic factors &16 &29  &20 &7 \\
	    \bottomrule
    \end{tabular}
    }
    \caption{How to set PerfConfs}
    \label{tab:howtoset}
\end{table}

\noindent \textbf{Format of PerfConfs}
A prior study shows configurations have many types \cite{xuTooManyKnobs}. PerfConfs, however, are dominated by numerical types. As shown in Table \ref{tab:howtoset},
the majority ($>$80\%) are integers, and a small
number of them ($\sim$10\%) are floating-point. There are 5 cases where the configurations are binary and determine whether a performance optimization is enabled.
A prior study shows that the difficulty of properly setting a configuration increases when the number of potential values increases \cite{xuTooManyKnobs}. Thus, due to their numeric types, PerfConfs are naturally difficult for users to set.

\noindent \textbf{Deciding Factors of PerfConf Setting}
We study what factors decide the proper setting of
a PerfConf based on developers' discussion and our source-code reading (Table \ref{tab:howtoset}).
In 2 cases, the setting depends only on static system features. For example, Cassandra suggests users set the \texttt{concurrent\_writes} to be $8\times number\_of\_cpu\_cores$. In 6 cases, it depends on workload features known before launch; e.g., input file size. Ideally, these PerfConfs would be set for each workload.

In most cases ($\sim$90\%), it depends on dynamic workload 
and environment characteristics, such as a job's dynamic memory consumption or node workload balance. For example, in CA6059, discusses \texttt{memtable\_total\_space\_in\_mb}, the maximum size of Cassandra server's in-memory write buffer. Depending on the workload's read/write ratio and the size of other heap objects, the optimal setting varies at run time. With no support for dynamic adjustment, Cassandra developers chose a conservative setting that lowers the possibility of OOM by sacrificing write performance for many workloads.


\subsection{Summary}


Our study shows that PerfConf problems are common in real-world software.  A single PerfConf often affects multiple performance metrics and its best setting may vary with workload and system. Thus, setting PerfConfs properly---i.e., to achieve the desired behavior in multiple metrics---is challenging for both developers and users. Ideally these software systems would  support users by automatically setting PerfConfs and dynamically adjusting them in response to changes in environment, workload, or users' goals.

\section{\tool Overview}
\label{sec:overview}

\myquote{``I don't know what idiot set this [configuration] to that.. oh wait, it was me...'' ---
 \hd{4618}}

\begin{figure}
\centering
\includegraphics[width=3.4in]{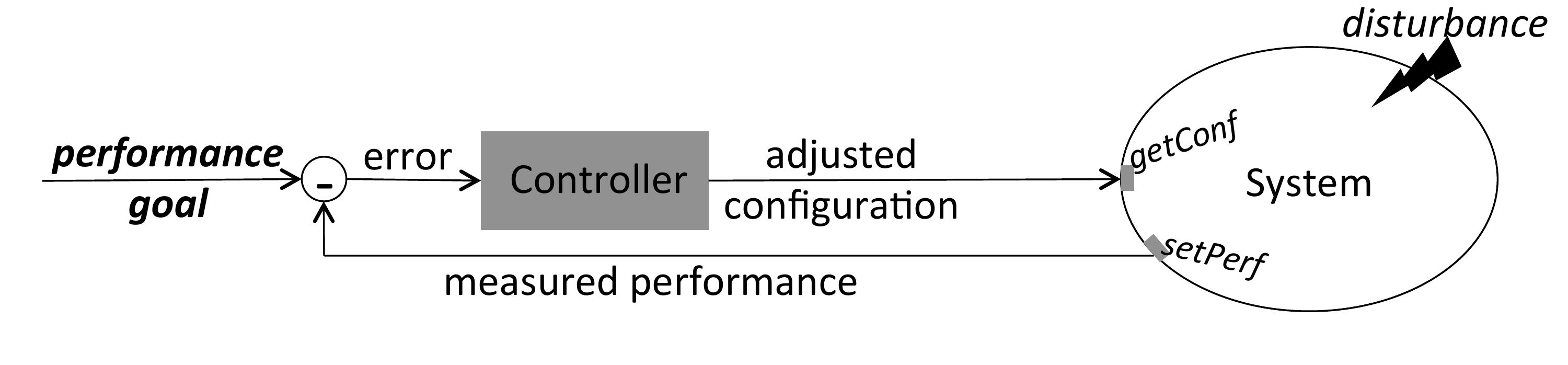}
\caption{Using a controller to adjust PerfConf (gray parts are extra controller-related components in \tool.)}
\label{fig:con}
\end{figure}

\vspace{-4mm}
\tool is a control-theoretic configuration framework for PerfConfs. 
As shown in Figure \ref{fig:con}, under \tool, users only need to specify
performance goals, instead of the exact configuration settings.
With small amount of refactoring, which we will detail later, the 
\tool-equipped system dynamically adjusts PerfConfs to satisfy user-defined performance goals---such as memory consumption constraints and tail latency requirements---despite unpredictable, dynamic environmental disturbances and workload 
fluctuations.

\paragraph{Why controllers?}
Machine learning (ML) and control theory are two options that can potentially
automate configuration. We choose control theory for
two reasons. 
First, controllers---unlike ML---are specifically designed to handle 
dynamic disturbances \cite{Hellerstein2004}, such as
environment changes and workload fluctuation, which is crucial
in setting many PerfConfs as discussed in Section \ref{sec:study_set}. 
A controller dynamically \textit{adjusts} a configuration based on the \textit{difference} between the current performance and the
goal, as illustrated in Figure \ref{fig:con}. In contrast, an ML model decides exact configuration
settings directly, which is more 
difficult in dynamically changing environments. 

Second, ML methods are better than controllers in deciding
optimal settings, which fortunately is unnecessary for most PerfConfs. As shown in Table \ref{tab:metricscount}, many configurations affect
memory and disk consumptions, where the main concern is not exceeding limits
instead of achieving a specific optimal value. Even for those PerfConfs
that affect request latencies, the corresponding goals are usually maintaining  
service-level-agreements, instead of achieving optimal latencies. Controllers are a good solution for meeting these types of constrained problems, because they provide formal guarantees that they will meet the constraint.  To handle PerfConfs, we modify standard control techniques, but still provide probabilistic guarantees.  ML would be a better choice if the goal was finding the best performance rather than meeting a performance constraint, and no guarantees were necessary.


\paragraph{What are the challenges?}
Our goal is to make control-theoretic benefits available to developers who are not trained in control engineering.
There are two high-level challenges: (1) how to automatically synthesize  controllers that can address unique challenges in the context of PerfConfs
and (2) how to allow developers to easily use controllers to adjust 
a wide variety of configurations in real-world software systems with little
extra coding.
We discuss how \tool addresses these two challenges in the next two sections.


\section{\tool Framework}
\label{sec:recipe}

\myquote{``It will be even great if we can dynamically tune/choose a proper one.'' --- \hb{7519}}


\tool provides a library for developers who want to have any configuration $C$ automatically and dynamically adjusted to meet 
a goal of a performance metric $M$, such as request latency, memory consumption, etc. 
This section describes what developers and users need to do to use \tool library and configurations.
Section \ref{sec:controller} describes how \tool library is implemented with new control theoretic techniques. 

\subsection{Developers' effort} 
\vspace{-2mm}
\subsubsection{General Code Refactoring}

First, developers must provide a sensor that measures the performance metric $M$ to be 
controlled.  Such sensors are sometimes already provided by existing software. 
For example, MapReduce contains sensors that measure and maintain up-to-date performance metrics in variables, such as heap consumption in \texttt{MemHeapUsedM}, average request latency in \texttt{RpcProcessingAvgTime}, etc.

Second, developers create a \tool system file invisible
to users, as shown in Figure \ref{fig:confemp}. In this system file, developers specify
the {mapping} from a \tool configuration entry $C$ to its corresponding 
performance metric $M$ and provide an initial setting for $C$, which only 
serves as $C$'s \textit{starting} value before the first run. After software starts, this field will be overwritten by the \tool controller. As we will see in our evaluation, the quality of this initial setting does not matter.

Third, developers replace the original configuration entry $C$ in the
configuration file with new entries 
$M.goal$ and $M.goal.hard$ 
that allow users to specify 
a numeric goal for $M$ and whether this goal is a hard constraint,
as shown in Figure \ref{fig:confemp}. 
For example, a goal about ``memory consumption should be smaller than the JVM heap size'' is a hard constraint.

\begin{figure}
  \lstset{basicstyle=\ttfamily\fontsize{8}{8}\selectfont,
     morekeywords={+},keepspaces=true,numbers=left}
  \mbox{
\begin{lstlisting}[mathescape,boxpos=t]
/* SmartConf.sys */
max.queue.size @ memory_consumption_max
max.queue.size = 50

/* HBase.conf */
memory_consumption_max = 1024
memory_consumption_max.hard = 1
\end{lstlisting}}
  \caption{\tool configurations}
  \label{fig:confemp}
\end{figure}

\subsubsection{Calling \tool APIs}
After the above code refactoring, developers can use \tool APIs.

\paragraph{Initializing a \tool Configuration}
Instead of reading a configuration value from the configuration file into an in-memory data structure,
developers simply create a \texttt{\tool} object \texttt{SC}. 
As shown in Figure \ref{fig:library}, the constructor's parameter is a string naming the configuration.
Using this string name, 
the \tool constructor reads the configuration's current setting,
its performance goal, and other \tool auto-generated parameters from the
\tool system file, and then initializes
a controller dedicated for this configuration, which we will explain more
in the next section.

\paragraph{Using a \tool Configuration}
Every time the software needs to read the configuration, \texttt{SC.setPerf} is invoked followed by
\texttt{SC.getConf}.
\texttt{setPerf} feeds the latest performance measurement \texttt{actual} to an underlying controller,
and \texttt{getConf} calls the controller to compute an adjusted configuration setting that can
close the gap between \texttt{actual} performance and the goal. 


\subsection{Handling Special Configuration Types}
The discussion above assumes a basic configuration that directly
affects performance all the time. 
Next, we discuss how \tool handles more complicated configurations. Only one type requires extra effort from developers.

\begin{figure}
\centering
  \lstset{basicstyle=\ttfamily\fontsize{8}{8}\selectfont,
     morekeywords={+},keepspaces=true,numbers=left}
  \mbox{
\begin{lstlisting}[mathescape,boxpos=t]
/* For direct configurations */
public class SmartConf{
  SmartConf (string ConfName); //initialize the controller
  void setPerf (double actual); //actual is obtained by a sensor
  int getConf (); //controller computes the adjusted setting
  void setGoal (double goal);
}
\end{lstlisting}}
  \caption{\tool class}
  \label{fig:library}
\end{figure}

\begin{figure}
  \centering
  \lstset{basicstyle=\ttfamily\fontsize{8}{8}\selectfont,
     morekeywords={+},keepspaces=true,numbers=left}
  \mbox{
\begin{lstlisting}[mathescape,boxpos=t]
/* For indirect configurations */
public class SmartConf_I extend SmartConf {
  SmartConf_I (string ConfName, Transducer t);
  void setPerf (double actual, int deputyConf);
}

/* Tranducer super class. Developers can customize a subclass.*/
public class Transducer {
  int transduce (int input) {return input};
}
\end{lstlisting}}
  \caption{\tool sub-class}
  \label{fig:libraryi}
\end{figure}

\paragraph{Indirect Configurations}
Sometimes, a configuration $C$ affects performance
indirectly by imposing constraints on its deputy $C'$. For example, in HBase,
\texttt{max.queue.size} limits the maximum size of a queue. The size of the queue, denoted as \texttt{queue.size}, then directly affects memory consumption. 
To handle indirect configurations like \texttt{max.queue.size}, a few steps in the above recipe need to change.

First, when creating the configuration object, 
developers should use the sub-type \texttt{SmartConf\_I} as shown in 
Figure \ref{fig:libraryi}. Furthermore, developers initialize the constructor with a \texttt{transducer} function that maps the desired value of deputy $C'$ to the desired value of configuration $C$. In most cases, this
transducer function simply conducts an identical mapping---if we want the
\texttt{queue.size} to drop to $K$, we drop
\texttt{max.queue.size} to $K$---and developers can
directly use the default transducer function provided by \tool library as
shown in Figure \ref{fig:libraryi}.

Second, while updating the current performance through 
\texttt{SC.setPerf}, developers need to provide 
the current value of $C'$---like the current \texttt{queue.size}, which is 
needed for the controller to adjust the value of $C$. The control theoretic reasoning behind this designed is explained in the
next section.

Finally, developers need to check every place where the configuration is
used to make sure that temporary inconsistency between the newly updated 
configuration $C$ and the deputy $C'$ is tolerated. 
For example, at run time the \texttt{queue.size} may be larger than a recently dropped \texttt{max.queue.size}. The right strategy is usually
to ignore any exception that might be thrown due to this inconsistency, and
simply wait for $C'$ to drop back in bound.
This change is needed for any system that supports dynamic configuration 
adjustment.
\vspace{-2mm}
\paragraph{Conditional Configurations}
As discussed in Section \ref{sec:study}, some configurations affect
 performance metrics conditionally. Consequently, the corresponding controller should only be invoked when the configuration takes effect.
Fortunately, this is already taken care of by the baseline \tool library, because developers naturally only invoke 
\texttt{SC.setPerf} and
\texttt{SC.getConf} when the software is to use the configuration.
\vspace{-2mm}
\paragraph{Correlating Configurations} Some configurations may
affect the same performance goal simultaneously, and their
corresponding controllers need to coordinate with each other.
This case is transparently handled by
\tool library and its underlying controllers synthesized by \tool.
As long as developers specify the same performance metric $M$ for a set of
configurations $\mathbb{C}$, \tool will make sure that their controllers
coordinate with each other. We will explain the control theoretic details in
the next section.
\vspace{-3mm}
\subsection{Users' Effort}
\vspace{-3mm}
With the above changes, users are completely relieved of directly setting
performance-related configurations.

In the configuration file, users simply provide two items to 
describe the
performance goal associated with a \tool configuration.
First, a numerical number that
specifies the performance goal, which could be the desired 
latency of user request, the maximum size of the memory consumption, etc. 
Second, a binary choice about whether or not the corresponding goal imposes a hard constraint. 
Developers provide default settings for these items, such as setting the memory-consumption goal to be the JVM heap size, just like that in traditional configuration files.
When users specify goals that cannot possibly be satisfied, \tool makes its best effort towards the goal and alerts users that the goal is unreachable. 

Users or administrators can update the goal at run time through the
\texttt{setGoal} API in Figure \ref{fig:library}.

\section{\tool Controller Design}
\label{sec:controller}

\myquote{``everything always has a tradeoff.'' --- \ca{13304}}

\vspace{-2mm}

\noindent \textbf{Baseline controller}
We choose a recently
proposed controller-synthesis methodology \cite{filieri2014automated} as the
foundation for the \tool controller.
This methodology first \textit{approximates} how system performance reacts to a configuration by profiling the application and building a regression model relating performance to configuration settings:
\vspace{-2.5mm}
\begin{equation}\label{eq:model}
\vspace{-0.1in}
s_k=\alpha \cdot c_{k-1}
\vspace{-0.03in}
\end{equation}
where $s_k$ is the system performance measured at time $k$ and $c_{k-1}$ is the 
configuration value at time $k-1$. A controller is then synthesized to  select the configuration parameter's next value $c_{k+1}$ based
on its previous value $c_{k}$ and the error $e_{k+1}$ between the desired
$\tilde s$ and measured performance $s_{k+1}$:
\vspace{-2.5mm}
\begin{equation}\label{eq:controller}
\vspace{-0.1in}
c_{k+1}=c_k+\frac{1-p}{\alpha}e_{k+1}.
\vspace{-0.03in}
\end{equation}
where $p$ is
the \emph{pole} value that determines how aggressively the controller reacts to
the current error. 

Although simple, the above controller is robust to model inaccuracy, and does not demand intensive profiling. We will explain more about this in Section \ref{sec:formal}.

\noindent \textbf{Challenges for \tool}
Unfortunately, the baseline controller, as well as \textit{all} existing control techniques, cannot handle several challenges unique and crucial to PerfConf problems.

\vspace{0.03in}
\noindent{1.} How to automatically set the pole $p$, to hide this control parameter from users. 

\vspace{0.03in}
\noindent{2.} How to handle hard goals that do not allow overshoot, such as memory consumption.

\vspace{0.03in}
\noindent{3.} How to handle the indirect relationships between some configurations and performance.

\vspace{0.03in}
\noindent{4.} How to handle multiple interacting configurations so that their controllers do not interfere with each other.
 
	
We explain how these challenges are addressed below.
	
\subsection{How to Decide the Pole Parameter}
\label{sec:controller_pole}

It is difficult for developers with no control background to set this value, so \tool sets it automatically. 

The pole $p$ determines the controller's tolerance for errors between the model built during profiling and the true behavior.  Given an error $\Delta$ between the true performance $s$ and the modeled performance $\hat{s}$, where $\Delta= s/\hat{s}$, the pole can simply be set to $p = 1 - 2/\Delta$, if $\Delta > 2$ and $p = 0$ otherwise.  Setting $p$ thusly guarantees the controller will converge \cite{Hellerstein2004}.

Of course, we do not expect \tool users to know $\Delta$, or even be aware of these control specific issues.
Therefore, \tool projects $\Delta$ based on the system's (in)stability during profiling:
$\Delta = 1+\frac{1}{N} \sum_{1}^{N}\frac{3\sigma_i}{{m_i}'}$, where $\sigma_i$ and ${m_i}’$ are the standard deviation and mean of the performance measured \wrt minimum performance under the $i$-th sampled configuration value. This equation provides a statistical guarantee that the controller will converge to the desired performance as long as the error between the model built during profiling and the true response is correct to within three standard deviations (i.e., 99.7\% of the time). 

\subsection{Handle Hard Goals}
\label{sec:controller_hard}
\vspace{-2mm}
Many PerfConfs are associated with a \texttt{hard==1} constraint, meaning that the goals like no OOM cannot be violated (Table \ref{tab:metricscount}). Handling these \emph{hard} constraints is crucial for system availability.
Unfortunately, traditional controllers can limit overshoot (i.e., the maximum amount by which the system may exceed the goal) only in continuous physical systems, \textbf{not} in discrete computing systems where a disturbance could come suddenly and discretely. For example, a new process could unexpectedly allocate a huge data structure.

\noindent \textbf{Strawman} 
One naive solution is to choose an extremely insensitive 
pole $p$ (e.g., close to 1), so that the output performance will move
very slowly towards the goal, making overshooting unlikely. Unfortunately, this
strategy does not work, as will be shown in experiments (Section
\ref{sec:eval}). It introduces extremely long convergence process, which
sacrifices other aspects of performance and still cannot prevent
overshooting when system dynamics encounter disturbance.

\noindent \textbf{A better strawman}
Recent work that uses controllers to avoid processor over-heating 
\cite{sironi2013thermos} proposes a \emph{virtual goal} $\tilde s^v$ that is smaller than the real constraint $\tilde s$.
The controller then targets $\tilde s^v$, instead of $\tilde s$. 
Unfortunately, this work still has two key limitations. First, while it works well for temperature---which changes slowly and continuously---it does not work well for goals like memory--which can change suddenly and dramatically. Second, it relies on expert knowledge to set the virtual goal $\tilde s^v$, without providing general setting methodology.

\noindent \textbf{Our Solution}
\tool proposes two new techniques to address goals that do not allow overshoot: automated virtual-goal setting and context-aware poles.

First, \tool proposes a general methodology to compute the virtual goal
$\tilde s^v$ considering system stability under control.
Intuitively, if the system is easily
perturbed, $\tilde s^v$ should be far from $\tilde s$ to
avoid accidental over-shooting. Otherwise, $\tilde s^v$ can be set to be
close to $\tilde s$ to allow better resource utility.  

To measure the system stability, we compute the coefficient of
variation $\lambda$
during the performance profiling at the model-building phase. 
That is, $\lambda := \frac{1}{N} \sum_{1}^{N}\frac{\sigma_i}{m_i}$, where $\sigma_i$ and $m_i$ are the standard deviation and mean of the 
performance measured under the $i$-th sampled configuration value. Clearly, the bigger $\lambda$ is, the more unstable the system is and hence the lower $\tilde s^v$ should be.
Consequently, we compute $\tilde s^v$ by (1-$\lambda$)*$\tilde s$. 

Second, 
\tool uses  
\textit{context-aware} poles that are conservative when the system is "safe" and aggressive when in ``danger''.
Specifically, before the virtual goal 
$\tilde s^v$ is reached, we use the regular pole, discussed in Section \ref{sec:controller_pole}. This 
pole is tuned to provide maximum stability given the natural system variance and may sacrifice reaction time for stability.
After $\tilde s^v$ is reached, we use the smallest possible pole, 0, which moves the system back in to the safe region as quickly as possible. 

As we can see, \tool handles hard goals without requiring any extra inputs from users or developers. The implementation of \tool API \texttt{SmartConf::getConf} will automatically switch to using the above algorithm (i.e., two poles and virtual goal) once the configuration file specifies a performance goal with the attribute \texttt{hard==1}. 
Experiments in Section \ref{sec:eval} demonstrate the above two techniques are crucial to avoid over-shooting while
maintaining high resource utility.

\subsection{Handle Configurations with Indirect Impact}
\label{sec:controller_indirect}
A PerfConf $C$ may serve as a threshold for a deputy variable $C'$  ($\sim$50\% among PerfConfs in Table \ref{tab:metricscount}). Directly modeling the relationship between performance and $C$ is difficult, as changing $C$ often does not immediately affect performance.

\tool handles this challenge by building a controller for the deputy variable $C'$ using the technique discussed earlier, and adjusting the threshold configuration $C$ based on the controller-desired value of $C'$.
Specifically, at run time, the controller computes the desired next value of $C'$ based on the current performance and the current value of $C'$, which is why the \texttt{SmartConf\_I::getPerf} function needs two parameters (Figure \ref{fig:libraryi}). \tool then adjusts $C$ to move $C'$ to the desired value. If $C$ simply specifies the upper-bound or lower-bound of $C'$, \tool sets $C$ to $C'_{\text{next}}$. 
If the relationship is more complicated, developers need to provide a custom \texttt{transducer} function as shown in Figure \ref{fig:libraryi}. 

For example, \tool profiles how software memory consumption changes with \texttt{queue.size}, and computes how to adjust \texttt{queue.size}
based on the current memory consumption.
If the desired size $q$ is smaller than 
\texttt{max.queue.size}, \tool drops \texttt{max.queue.size} to $q$. 
This does not immediately shrink \texttt{queue.size}, but will prevent the queue from taking in new RPC requests until \texttt{queue.size} drops into the
desired range. 


\subsection{Handle Multiple, Interacting PerfConfs}
\label{sec:controller_mul}

The discussion so far assumes \tool creates an independent controller for each individual configuration. It is possible that multiple configurations---and hence multiple controllers---are associated with the same performance constraint, as implied by Table \ref{tab:metricscount}. We must ensure that each controller works with others towards the same goal. For example, when two controllers independently decide to increase \texttt{q1.size} and \texttt{q2.size}, \tool must ensure no OOM.

Traditional control techniques synthesize a single controller that sets all configurations simultaneously. This approach demands much more complicated profiling and controller building, essentially turning a O($K\cdot N$) problem into a $O(N^K)$ problem, assuming $K$ PerfConfs each with $N$ possible settings \cite{filieri2015multi}. Furthermore, it is fundamentally unsuitable for PerfConfs, as different PerfConfs may be developed at different times as software evolves, and they may be used in different modules and moments during execution. We assume developers will call \texttt{getPerf} and \texttt{setConf} at the places the program uses a PerfConf value.  Traditional techniques for coordinating control would require all \texttt{getPerf} 
and \texttt{setConf} calls be made in the same location at the same time, which we believe is infeasible in a large software system. 

Therefore, instead of synthesizing a single controller to set all configurations simultaneously, \tool uses a protocol 
such that controllers will independently work together. When we synthesize the controller for $C$, the performance impact of related configurations is part of the disturbance captured during profiling and hence affects how \tool determines the pole 
(Section \ref{sec:controller_pole}) and the virtual goal (Section \ref{sec:controller_hard}). As we will discuss soon in Section \ref{sec:formal}, even if the profiling is incomplete, our controller-synthesis technique still provides statistical guarantees that the goal will be satisfied.

When developers are extremely cautious about not violating a performance goal or feel particularly unsure about the profiling, \tool provides a safety net by applying an interaction factor $N$ to Equation 2. Specifically, developers can mark a specific performance goal---e.g., memory consumption or 99 percentile read latency---as \textit{super}-hard. While processing the \tool system file, \tool counts how many, configurations are associated with this super-hard goal. Then, when initializing a corresponding controller $c$, \tool will use 
$c_k + \frac{1-p}{N\alpha}e_{k+1}$ instead of
$c_k + \frac{1-p}{\alpha}e_{k+1}$ as the formula to compute the setting of $c_{k+1}$, splitting the performance gap $e_{k+1}$ evenly to all $N$ interacting configurations.

\subsection{Other Implementation Details}
\label{sec:impl}
Our \tool library is implemented in Java. 
The \texttt{SmartConf} classes shown in Figure \ref{fig:library} and Figure \ref{fig:libraryi} contain private fields representing the configuration name \texttt{ConfName}, current configuration setting, current performance, and controller parameters, including pole, $\alpha$, goal, and virtual goal (for \texttt{SmartConf\_I} class).
These controller parameters are computed inside the \texttt{SmartConf} constructor based on the profiling results stored in a configuration-specific file \texttt{<ConfName>.SmartConf.sys}. Of course, future implementations can change to compute these parameters only once after all the profiling is done. 

\begin{table*}[t]

{\small
\begin{tabular}{l@{\hspace{0.05in}}l|l|ll}
\toprule
   &                   &Profiling& \multicolumn{2}{c}{Evaluation Workload}\\
   \cline{4-5}
ID & Issue Description &Workload & Phase-1 & Phase-2\\
\midrule
\href{https://issues.apache.org/jira/browse/CASSANDRA-6059}{CA6059} &\texttt{memtable\_total\_space\_in\_mb} limits the memtable size.&\multirow{7}{*}{YCSB$_A$} 			& YCSB						& YCSB\\
N-N-Y     &Too big, OOM; Too small, write latency hurts.					 &\multirow{8}{*}{\scriptsize{0.5W, 1MB}}&{\scriptsize{1.0W, 1MB, C0}} &{\scriptsize{0.9W, 1MB, C0.5}}\\
\cline{1-2}\cline{4-5}
\href{https://issues.apache.org/jira/browse/HBASE-2149}{HB2149}	&\texttt{global.memstore.lowerLimit} decides how much memstore data is flushed.&&YCSB					&YCSB					\\
Y-Y-N     &Too big, write blocked for too long;  Too small, write blocked too often.	   &&\scriptsize{1.0W, 1MB, 10s} &\scriptsize{1.0W, 1MB, 5s}\\
\cline{1-2}\cline{4-5}
\href{https://issues.apache.org/jira/browse/HBASE-3813}{HB3813}		 &\texttt{ipc.server.max.queue.size} limits RPC-call queue size.				&&YCSB					&YCSB\\
N-N-Y      &Too big, OOM; Too small, read/write throughput hurts.								&&\scriptsize{1.0W, 1MB} &\scriptsize{1.0W, 2MB}\\
\cline{1-2}\cline{4-5}
\href{https://issues.apache.org/jira/browse/HBASE-6728}{HB6728}	 &\texttt{ipc.server.response.queue.maxsize} limits RPC-response queue size.&&YCSB					&YCSB\\
N-N-Y		&Too big, OOM; Too small, read/write throughput hurts.									&&\scriptsize{0.0W, 2MB} &\scriptsize{0.3W, 2MB}\\
\midrule
\href{https://issues.apache.org/jira/browse/HDFS-4995}{HD4995} &\texttt{content-summary.limit} limits \#files traversed before \texttt{du} releases big lock.&TestDFSIO&TestDFSIO&TestDFSIO\\
Y-N-N     &Too big, write blocked for long; Too small, \texttt{du} latency hurts.						&\scriptsize{single-thread}&\scriptsize{multi-thread, 20s}&\scriptsize{multi-thread, 10s}\\
\midrule
\href{https://issues.apache.org/jira/browse/MAPREDUCE-2820}{MR2820}  & \texttt{local.dir.minspacestart} decides if a worker has enough disk to run task.&WordCount&WordCount&WordCount\\
Y-Y-Y     & Too small, OOD; Too big, low utility (job latency hurts).	&\scriptsize{2G, 64MB, 1}&\scriptsize{640MB, 64MB, 2}&\scriptsize{640MB, 128MB, 2}\\
\bottomrule
\end{tabular}
}
\caption{Benchmark suite and workload. \textmd{?-?-? under a bug ID shows whether the PerfConf is conditional, direct, and hard. In issue description, the main constraint that users complain about is put earlier, and the trade-off is later. For YCSB \cite{cooper2010benchmarking} workload, \texttt{x}W, write portion; \texttt{y}MB, request size; C\texttt{z}, read index cache ratio. \texttt{t}s, latency constraint. Wordcount(x,y,z): input file size; split size; parallelism per worker}}
\label{tab:evalbench}
\end{table*}

The \tool system file \texttt{SmartConf.sys} contains an entry that allows developers to enable or disable profiling. Once profiling is enabled, the calling of \texttt{SmartConf::setPerf}  records the current performance measurement not only in the \texttt{SmartConf} object but also in a buffer, together with the current (deputy) configuration value, periodically flushed to file  \texttt{<ConfName>.SmartConf.sys}, which 
will be read during the initialization of configuration \texttt{<ConfName>}.

\noindent \textbf{Profiling}
To model the effects the controller has on the target performance metric, a few performance measurements need to be taken by running profiling workloads while varying the configuration parameter to be controlled.
The larger the range of workloads, the more robust the control design will be when working with previously unseen workloads. 
We also base the pole and the virtual goal on the measured mean and standard deviation, so enough samples are needed for the central limit theorem to apply.
As we will formally discuss below and experimentally demonstrate in Section \ref{sec:eval}, \tool produces effective and robust controllers without intensive profiling.

\subsection{Formal Assessment and Discussion}
\label{sec:formal} 

\vspace{-3mm}
\noindent \textbf{Stability}
We want the system under control to be 
\textit{stable}. That is, it should converge to the desired goal rather than oscillate around it, which could cause unpredictable performance or crashes.
Based on analysis in previous work the controller in equation \ref{eq:controller} is stable as long as $0 \le p < 1$ and $p = 1 - 2/\Delta$ for $\Delta > 2$ \cite{filieri2014automated}.  Unlike prior work, \tool assumes $\Delta$ is unknown, so we provide a weaker probabilistic guarantee that the system will converge as long as the error is within three standard deviations of the true value. This guarantee comes without requiring users to have control-specific knowledge.

\noindent \textbf{Overshoot}
We hope to ensure that hard goals that do not allow overshoot are respected. Following traditional control analysis \tool is free of overshooting because its design ensures $0 \le p < 1$ \cite{Hellerstein2004}. Such analysis, however, assumes no disturbances, but we know we are deploying \tool into unpredictable environments.

With two enhancements discussed in Section \ref{sec:controller_hard}, we avoid overshooting with high probability even in unpredictable environments.  By setting the virtual goal to  $\lambda := \frac{1}{N} \sum_{1}^{N}\frac{\sigma_i}{m_i}$, we provide 84\% probability of being on the "safe" side of no-overshoot goals.\footnote{Assuming a normal distribution, 68\% of samples are within 1 standard deviation, which means 16\% is higher and 16\%.  In our case, however, one side is safe, so we have an 84\% probability of not overshooting.} The two-pole enhancement further increases the likelihood that \tool respects the constraint, because any measurement above the virtual goal causes the largest possible reaction in the opposite direction.

\section{Evaluation} 
\label{sec:eval} 

\myquote{``I think going to 1G [default] works, ... let's do some testing before submitting a patch'' -- \hb{4374}}
\subsection{Evaluation methodology}
\label{sec:eval_meth}
\vspace{-2mm}
\noindent \textbf{Benchmarks}
We apply \tool to 6 PerfConf issues in Cassandra, HBase, HDFS, and MapReduce, as shown in Table \ref{tab:evalbench}. These 6 cases together cover a variety of configuration features, like conditional or not, direct or not, hard constraint or not, as listed by ?-?-? sequence in Table \ref{tab:evalbench}. We consider bug-reporters' main concern as the performance goal, and the trade-off mentioned by users or developers as the trade-off metric that we want to optimize while satisfying the goal, both listed in the issue description of Table \ref{tab:evalbench}. They cover a variety of performance metrics, memory, disk, latency, etc. 

\noindent \textbf{Workloads}
\tool works in a wide variety of workload settings, but we do not have space to show that. Therefore, in this section, our workload design follows several principles: (1) profiling and evaluation workload are \textbf{different}, so that we can evaluate how sensitive \tool is towards profiling; 
(2) the evaluation workload contains two phases where either the workload or the performance goal changes (HB2149, HB6728), so that we can evaluate how well \tool reacts to changing \textbf{dynamics}; (3) at least one phase of the evaluation workload triggers the performance \textbf{problems} complained by users in the original bug reports, so that we can test whether \tool automatically addresses users' PerfConf problems. Finally, we use standard profiling workloads to demonstrate \tool's robustness. Specifically, for key-value stores, we use the popular YCSB \cite{cooper2010benchmarking} benchmark workload-A, which has a 50-50 read-write ratio; for HDFS, we use a common distributed file system benchmark TestDFSIO \cite{huang2010hibench}; and we use WordCount for MapReduce, as shown in Table \ref{tab:evalbench}.

\noindent \textbf{Machines}
We use two servers to host virtual machines. Each server has 2 12-core Intel Xeon E2650 v3 CPU with 256GB RAM. Ubuntu 14.04 and JVM 1.7 are installed. We use virtual machines to host distributed systems under evaluation, with 2--6 virtual nodes set up for each experiment.

\subsection{Does \tool Satisfy  Constraints?}
\vspace{-2mm}
\tool always tracks the changing dynamics, satisfying the performance constraints for all 6 issues. These  include hard constraints---preventing out-of-memory (CA6059, HB3813, HB6728) and out-of-disk (MR2820) ---and soft constraints on worst-case write latency (HB2149, HD4995). 

\begin{figure}
\centering
\includegraphics[width=3.3in]{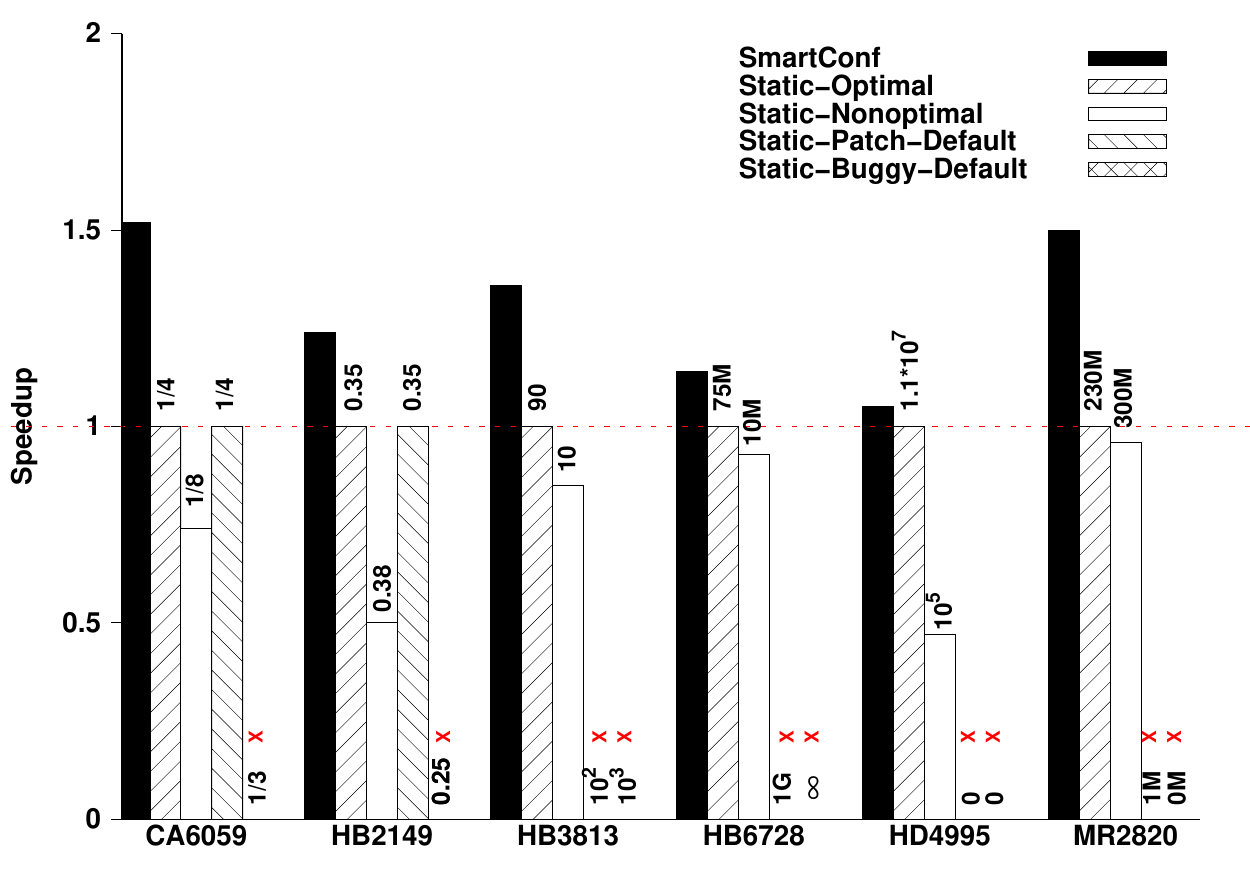}
\caption{Trade-off performance comparison. \textmd{Normalized upon the best-performing static configuration;  {\color{red}x}: fail the perf. constraint. The numerical PerfConf settings are above each bar.}}
\label{fig:overallresult}
\end{figure}

It is difficult for statically set configurations to satisfy performance constraints. The original default settings in all 6 issues fail, denoted by the red-crosses for \texttt{static-buggy-default} bars in Figure \ref{fig:overallresult}, which is why users filed issue reports. In our experiments, even the patched default settings fail to satisfy corresponding constraints in 4 cases. In HD4995, developers simply moved a problematic hard-coded parameter into the configuration file without changing the default setting, and asked users to figure out a suitable custom setting for themselves. In HB3813, HB6728, and MR2820, the patches made the configurations more conservative, from 1000, $\infty$, and 0 to 100, 1$G$, and 1$M$ respectively. However, the new settings still failed. In fact, we can easily find workloads to make the patched default settings in the remaining 2 issues fail, too.

\noindent \textbf{Case Study} We take a closer look at how \tool handles HB3813. Here, \texttt{max.queue.size} decides the largest size for an RPC queue. When the system is under memory pressure, a large queue can cause an out-of-memory (OOM) failure. Unfortunately, a small queue reduces RPC throughput.

\begin{figure}
\centering
\includegraphics[width=3.5in]{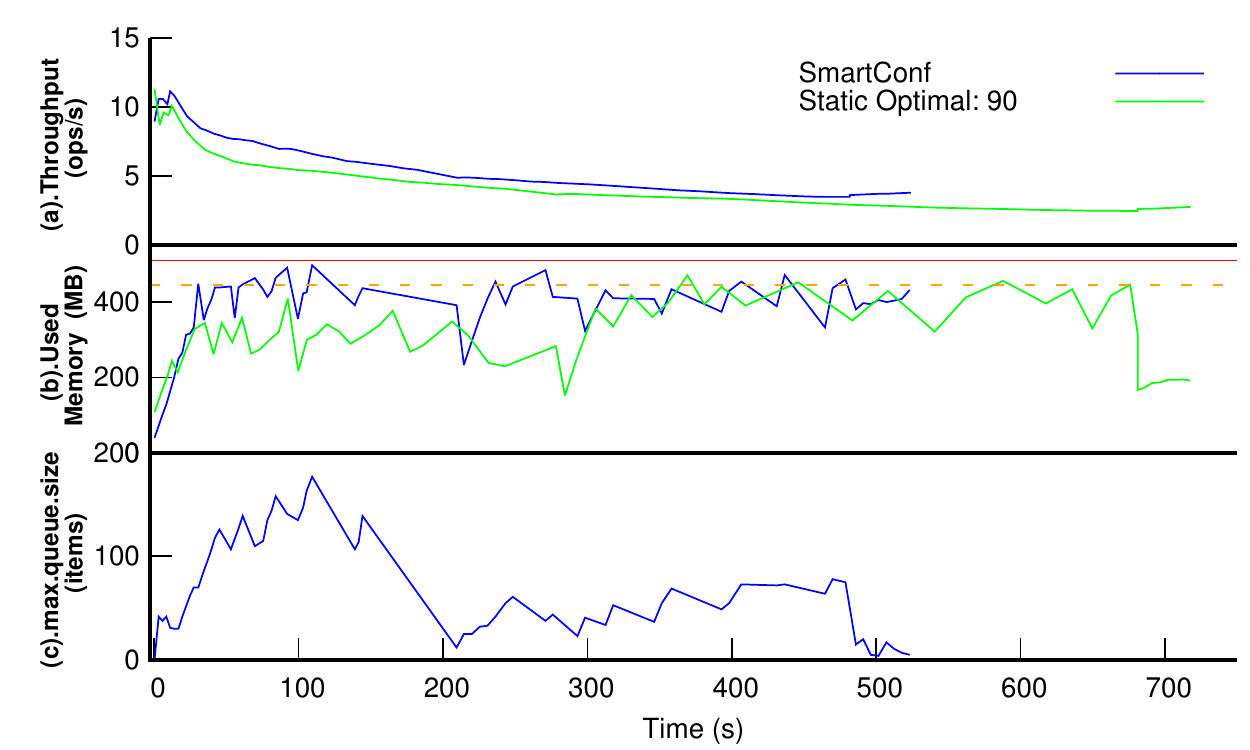}
\caption{\tool vs. static optimal on HB3813. \textmd{workload changes at $\sim$200s. Throughput is accumulative.}}
\label{fig:hbase3813basic}
\end{figure}

Figure \ref{fig:hbase3813basic}b shows how memory consumption changes at run time under different configuration settings. The red horizontal line marks the hard memory-consumption constraint (495MB), and the orange dashed line marks \tool's automatically determined virtual goal of 445MB. The blue curve shows how memory consumption changes under \tool's automated management. While under the dashed line---in a "safe zone"---the system takes new RPC requests, \tool slowly raises \texttt{max.queue.size} from its initial value $0$---shown by the blue curves in Figure \ref{fig:hbase3813basic}c---and the memory consumption increases. Once over the dashed line, \tool quickly decreases \texttt{max.queue.size}---shown by the dips of the blue curve in Figure \ref{fig:hbase3813basic}c---and the memory drops. 
Even when the workload shift increases each RPC request size (at about the 200 second point), the memory consumption is always under control, as \tool reacts to the workload change by dropping the 
\texttt{max.queue.size} to around 50---as shown by the blue curves---after 200 seconds in Figure \ref{fig:hbase3813basic}c. Overall, the system never has OOM errors with \tool.

In comparison, the old default \texttt{max.queue.size} setting, 1000, causes OOM almost immediately after the first workload starts; even the new default setting in the patch, 100, still causes OOM shortly after the second workload starts.
A conservative setting---e.g., 90 in this experiment---avoids OOM, shown by the green curves in Figure \ref{fig:hbase3813basic}ab. However, there is no way for users or developers to predict what configuration will be conservative enough for future workload.

\subsection{Does \tool Provide Good Tradeoffs?}
\vspace{-2mm}
Figure \ref{fig:overallresult} shows that \tool provides performance tradeoffs better than the best static configuration.  While all of our case studies have different constraints, they all must optimize latency or throughput under those constraints.  The figure shows \tool's speedup in these secondary metrics relative to various static configurations.

We find the best static configuration by exhaustively searching all possible PerfConf settings that meet the constraint throughout our two-phase workloads. These best settings are often sub-optimal or even fail the performance constraints once workloads change. Figure \ref{fig:overallresult} also shows the performance under non-optimal static settings that we randomly choose. 

\tool outperforms the best static setting because it automatically adapts to dynamics. 
Although \tool may start with a poor initial configuration (e.g., 0 in Figure \ref{fig:hbase3813basic}c), it quickly adjusts so that the constraint is \textit{just} met and the tradeoffs are optimal. When the workload changes from phase-1 to phase-2 in our experiments, \tool quickly adjusts again. In comparison, since different phases have different constraints, a static configuration can only be optimal for one phase and must sacrifice performance for the other .

For example, as shown in Figure \ref{fig:hbase3813basic}ab, to avoid OOM during both phases, the static optimal configuration (90) is too conservative and unnecessarily reduces memory during the first phase. In contrast, \tool is never too conservative or too aggressive. Throughout the two phases, \tool achieves 1.36$\times$ speedup in write throughput.

As another example, in MR2820, to make sure WordCount can succeed in both phases, the best static setting for
\texttt{minspacestart} is 230MB, because phase-2 requires that much disk space to run. However, this is overly conservative for phase-1 that produces much smaller intermediate files. Consequently, \tool runs WordCount much faster in phase-1, and achieves 1.50$\times$ total speedup.

\subsection{Alternative Design Choices} 
\vspace{-2mm}  
\tool's controller handles hard constraints differently from traditional control design in two ways (Section \ref{sec:controller_hard}).  
We experimentally compare with the traditional alternatives below.

\begin{figure}
\centering

\centering
\includegraphics[width=3.3in]{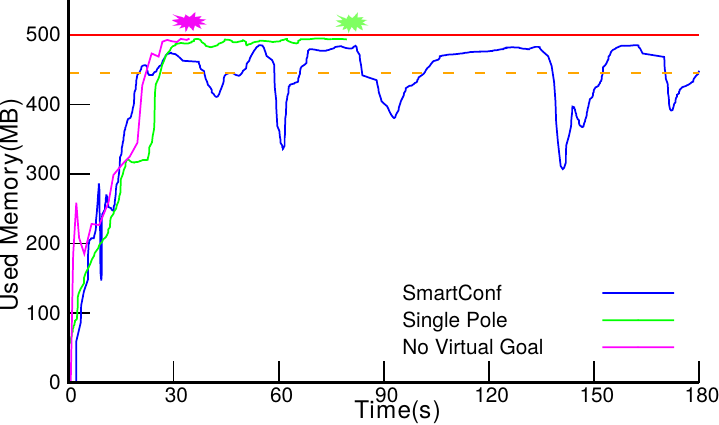}
\caption{\tool vs. alternative controllers.}
\label{fig:alternative}
\end{figure}

\noindent \textbf{A Single Pole with a Good Virtual Goal} Traditional control design handles hard constraints---e.g., avoiding processor over-heating \cite{sironi2013thermos}---by using a single conservative pole and a virtual goal.  We briefly compare this traditional design to \tool by recreating the HB-3813 case study using a less stable workload (70\% write with 30\% read). We let \tool and this alternative controller use the same virtual goal and the same pole 0.9. The only difference is that \tool has a second pole, 0, for post-virtual-goal use.

As shown in Figure \ref{fig:alternative}, \tool still behaves well, yet the single-pole alternative controller causes an OOM at time 80s.
Around 25s, both controllers start to limit queue size, but the alternative one is simply too slow. When close to the memory limit---i.e., beyond the virtual goal---that slowness is catastrophic because just a few extra RPC requests can cause a system crash.  
Overall, \tool is conservative when growing the queue and extremely aggressive when shrinking it. In contrast, with only one pole, the alternative controller is conservative when growing the queue and too conservative when shrinking it.  

\noindent \textbf{Without (a good) Virtual Goal}
Traditional control design does not consider virtual goals.  We rerun the HB3813 example, this time targeting the actual system memory instead of the virtual goal determined by \tool. As shown in Figure \ref{fig:alternative}, the system quickly over-allocates memory leading to a JVM crash at about 36s.  The virtual goal is essential for meeting hard constraints because it gives \tool's controller time to react to unexpected situations. Needless to say, selecting the right virtual goal is crucial. A careless selection easily leads to violating constraints or wasting resources. We skip the experimental results here due to space constraints.

\label{sec:others}
\begin{figure}
\centering
\includegraphics[width=3.3in]{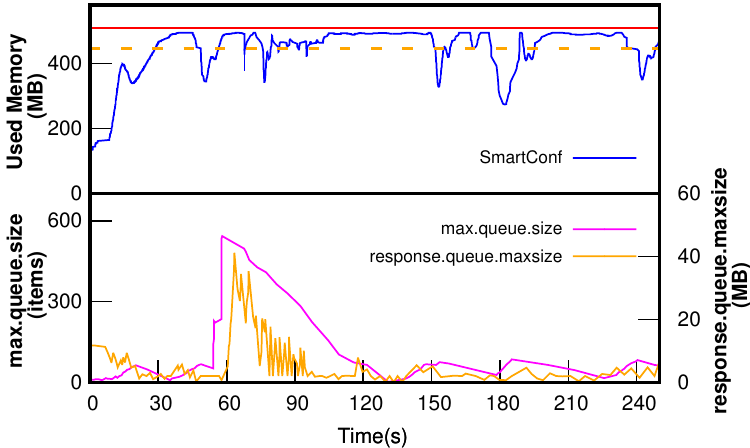}
\caption{\tool adjusts two related PerfConfs.}
\label{fig:hbasecombine}
\end{figure}

\begin{table}
\centering
{\small
\begin{tabular}{lrrrr}
	\toprule
    ID & Sensor & Invoke APIs & Others&Total\\
    \midrule
    CA6059& 35  &6&1&42\\
    HB2149&31   &6&1&38\\ 
    HB3813& 2  &6&9&17\\
    HB6728& 2  &6&0&8\\
    HD4995& 70 &6&0&76\\
    MR2820& 53 &8 &4&65\\
    \bottomrule
    \end{tabular}
    }
    \caption{Lines of code changes for using \tool}
    \label{tab:refactor}
    \end{table}

\subsection{Other results}
\noindent \textbf{Is \tool easy to use?}
As shown in Table \ref{tab:refactor}, it takes little effort for developers, as few as 8 lines of code changes. The  changes are dominated by implementing performance sensing. For HB-3813, extra changes are needed to tolerate the temporary inconsistency between a deputy variable (\texttt{queue.size}) and the indirect configuration (\texttt{max.queue.size}). For MR-2820, extra changes are needed to deliver the configuration computed by one node (Master) to another (Slave).

\noindent \textbf{Interacting controllers}
To evaluate whether \tool can handle multiple interacting PerfConfs, as mentioned in Section \ref{sec:controller_mul}, we apply \tool to tackle HB3813 and HB6728 simultaneously. The PerfConfs in these two cases limit the size of RPC-request queue and RPC-response queue, respectively, both affecting memory consumption. We start with a workload of writes, occupying a large space in the request queue and a small space in the response queue. After 50 seconds, we add a second workload of reads, which take small space in the request queue and large space in the response queue. 
Figure \ref{fig:hbasecombine} shows the results. When the second workload just starts, the request queue quickly fills with many small read requests, and the response queue jumps up. Then, \tool reacts by bringing the size of both queues down dramatically.  After this initial disturbance, the size of each queue dynamically fluctuates: during periods where more read requests enter the system, the response queue size is limited; when there are more write request, the RPC queue size is throttled.  At no time is the memory constraint (red line) violated.  

This study demonstrates that multiple PerfConfs can be composed and still guarantee the hard constraint.  It also further motivates dynamically adjusting configurations: otherwise, we would have to pick very small sizes for both queues.

\subsection{Limitations of \tool}
\tool also has its limitations. First, it does not work for configurations whose performance goals are about
optimality instead of constrained optimality.
For example, MR-5420 discusses how to set \texttt{max\_chunks\_tolerable} which decides how many chunks that input files can be grouped into
during distributed copy. The on-line discussion shows that users only care about one goal here --- achieving the fastest copy speed. Consequently, \tool is not a good fit. Second, the current \tool design does not work if the relationship between performance and configuration is not monotonic. This happens to be the case in MR-5420 --- if there are too few chunks, the copy is slow due to load imbalance; if there are too many chunks, the copy is also slow due to lack of batching. Machine learning techniques would be a better fit for these two challenges. Third, some configurations might be inherently difficult to adjust dynamically, as the adjustment may be expensive. For example, changing \texttt{max\_chunks\_tolerable} dynamically may require copying files around. Finally, \tool provides statistical guarantees as discussed in Section \ref{sec:formal}, but cannot guarantee all constraints to be always satisfied.




\section{Related Work} 
\label{sec:related}  
\myquote{``What is the proper way of setting the configuration values programmatically?'' -- \href{https://stackoverflow.com/questions/12825547/setting-hadoop-job-configuration-programmatically}{MapReduce-12825547}
}
\vspace{-4mm}
\paragraph{Control theoretic frameworks}
Control theory provides a general set of mechanisms and formalisms for ensuring that  systems achieve desired behavior in dynamic environments \cite{levine1996control}.  While the great body of control development has targeted management of physical systems (e.g., airplanes), computer systems are natural targets for control since they must ensure certain behavior despite highly dynamic fluctuations in available resources and workload \cite{Karamanolis,hellerstein2004challenges}.  

While control theory covers a wide variety of general techniques, control applications tend to be highly specific to the system under control.  The application-specific nature of control solutions means that controllers that work well for one system (e.g. a web-server \cite{LuEtAl-2006a,SunDaiPan-2008a} or mobile system \cite{Agilos}) are useless for other systems.  

Thus, a major thrust of applying control theory to computing systems is creating general and reusable techniques that put control systems in the hands of non-experts \cite{ControlSurvey}.  Towards this end, recent research synthesizes controllers for software systems \cite{filieri2014automated,filieri2015multi,shevtsov2016}.  Other approaches package control systems as libraries that can be called from existing software \cite{ControlWare,swift,POET}. While these techniques automate much of the control design process, they still require users to have control specific knowledge to specify key parameters, like the values of $p$ and $\alpha$, and choose what  controllers to use.  Furthermore, none of them address the PerfConf specific challenges of meeting hard constraints, using indirect and interacting parameters, etc.    

In addition to solving PerfConf-specific challenges, \tool is unique in hiding all control-specific information from the users/developers.  Thus, \tool's interface works at a much higher-level of abstraction than prior work that encapsulates control systems.  In fact, \tool's implementation could swap a control system for some other management technique in the future. In exchange for its higher level of abstraction, \tool provides only probabilistic guarantees rather than the stronger guarantees that would come from having an expert set a pole based on a known error bound.

Many learning approaches have been proposed for predicting an optimal configuration within a complicated configuration space \cite{starfish.cidr11,morpheus.osdi16,pavlo.sigmod17}.  
Perhaps the most closely related learning works are those based on reinforcement learning (RL) \cite{RL}. Like control systems RL takes online feedback.  Several RL methods exist for optimizing system resource usage \cite{ganapathi2009case,Ipek2006,Ipek2008,Bitirgen2008,Martinez2009}.  RL techniques, however, are not suited to meeting constraints in dynamic environments \cite{RL-book}.  In contrast, that is exactly what control systems are designed to do, and they produce better empirical results than RL on such constrained optimization problems \cite{TAAS}. 

\noindent \textbf{Misconfiguration}
Many empirical studies have looked at misconfiguration \cite{fragile,ricardo.osdi04,xuTooManyKnobs,zuoning.sosp11}, but did not focus on PerfConfs.
Much previous work has proposed using static program analysis \cite{katz.ase11, pcheck.osdi16} or statistical analysis \cite{strider04, wang-peerpressure,ding.atc11,encore.asplos14} to identify and fix wrong or
abnormal configurations. These techniques mainly target functionality-related misconfigurations, and do not work for PerfConfs, as the proper setting of a PerfConf highly depends on the dynamic workload and environment, and can hardly be statistically decided based on common/default settings.
Techniques were also proposed to diagnose misconfiguration failures \cite{confaid.osdi10,YiMingTraceToolOSDI06} and misconfiguration-related performance problems \cite{xray.osdi12}. They are complementary to \tool that helps avoid misconfiguration performance problems.

\section{Conclusions}
\vspace{-2mm}
Large systems are often equipped with many configurations that allow customization. Many of these configurations can greatly affect performance, and their proper setting unfortunately depends on complicated and changing workloads and environments. We argue that the traditional way of letting users statically and directly set configuration values is fundamentally flawed. Instead, a new configuration interface is designed to allow users and developers to focus on specifying what performance constraints a configuration should satisfy, and a control-theoretic technique is designed to enable automated and dynamic configuration adjustment based on the performance constraints. Our evaluation shows that the \tool framework can often out-perform static optimal configuration setting, while satisfying performance constraints.
\newpage

\newpage
\bibliographystyle{plain}
\bibliography{references,shan}

\end{document}